# Anomalous Orbital Magnetism and Hall Effect of Massless Fermions in Two Dimension

Hidetoshi FUKUYAMA[1]*

[1]*Department of Applied Physics, Faculty of Science, Tokyo University of Science, 1-3 Kagurazaka, Shinjuku-ku, Tokyo 162-8601*

Inter-band effects of magnetic field on orbital magnetic susceptibility and Hall effect in weak magnetic field have been studied theoretically at absolute zero for the model of massless Fermions in two dimension described by Weyl equation similar to graphenes, which are simplified version of newly-found one in molecular solids, $\alpha$-ET$_2$I$_3$, described by the "tilted Weyl" equation. The dependences on the Fermi energy of both orbital susceptibility and Hall conductivity near the zero-gap region scale with the elastic scattering time and then are very singular.

KEYWORDS: massless Fermions, graphene, $\alpha$-ET$_2$I$_3$, Hall effect, orbital magnetism, tilted Weyl equation

Effects of magnetic field in solids are very intricate, since magnetic fields (or vector potential) have matrix elements between Bloch bands, and have been known as "inter-band effects".[1] The typical example is the orbital magnetism. Actually the orbital magnetic susceptibility in weak magnetic fields has many contributions whose classification is not simple in contrast to the clear Landau diamagnetism of free electrons.[2] The simplest case of important inter-band contributions is the atomic diamagnetism which is present even if there is no finite density of states (DOS) at the Fermi energy (i.e. even in the atomic limit and then in insulators). The existence of finite orbital magnetism without any DOS at the Fermi energy is in sharp contrast to the Landau-Peierls (LP) formula, which is exact for free electrons but approximate for Bloch electrons. The LP formula is derived by use of the Peierls phase due to vector potential within a particular band, and then is often termed as "one-band formula". The validity of this formula has been questioned even from early days.[3] Actually it turns out that LP formula is not valid even in the case where the periodic potential is very weak.[4] Experimentally, anomalous magnetism has been observed in graphite[5] and bismuth and Bi-Sb alloys[6,7] both of which are narrow gap semimetals. These experimental facts contradict with LP formula in an obvious way, i.e. the experiments indicate that the diamagnetism has largest values when the density of states at the Fermi energy is vanishing or very small. These phenomena have been theoretically understood as due to inter-band contributions with different degree of importance of spin-orbit interaction, expressed by the model Hamiltonians whose

---

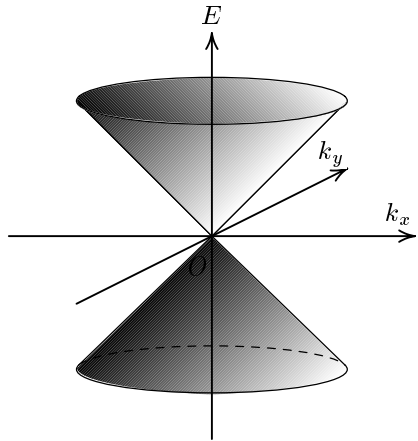

Fig. 1. The energy spectrum of massless Fermions in two-dimension.

essential features are described by the Weyl equation for neutrinos in the case of graphite[8] and the Dirac equation in the case of bismuth.[9]

Together with such orbital magnetism, Hall effects in the weak magnetic field has been a long-standing problem[1,10,11] and not so much is known so far in contrast to that in strong magnetic field, i.e. quantum Hall effect.[12]

Because of recent experimental realizations of similar but more dramatic situations with truly zero-gap state ("massless fermions") in graphene[13,14] and $\alpha$-ET$_2$I$_3$,[15] there are naturally renewed interests in the inter-band contributions in orbital susceptibility and Hall effects, which will be studied theoretically in this paper at absolute zero temperature based on the general formulas of orbital susceptibility[16] and Hall conductivity.[17] We note that the Hall conductivity of graphene in the presence of strong magnetic field have been studied by Zheng and Ando[18] for the case of strong magnetic field but the case of weak filed limit has not been studied in detail because of the singular behaviors near the zero-gap region. We take unit of $\hbar = k_B = 1$.

The model we will study is two dimensional electrons whose energy spectra in the $x$-$y$ plane is given by 2×2 matrix as follows in the Luttinger-Kohn(LK) representation[19]

$$\mathcal{H} = v \begin{bmatrix} 0 & k_x - ik_y \\ k_x + ik_y & 0 \end{bmatrix} = v(k_x \sigma_x + k_y \sigma_y) \qquad (1)$$

where $v$ is the velocity and $\sigma_\alpha$ ($\alpha = x, y$) is the Pauli matrix. The energy spectra $E(\boldsymbol{k})$, i.e. Bloch bands, is seen to be $E(\boldsymbol{k}) = \pm vk$ ($k = |\boldsymbol{k}|$) by diagonalization of the matrix and shown in Fig. 1. Here the origin of energy is taken at the crossing point. In the presence of magnetic field applied perpendicular to the two-dimensional plane, $\boldsymbol{k}$ is replaced by $\boldsymbol{\pi} = \boldsymbol{k} + e\boldsymbol{A}/c$ with $-e$ and $c$ being the electronic charge and the light velocity, respectively, and $\boldsymbol{A}$ is vector potential associated with magnetic field $\boldsymbol{H}$ by $\boldsymbol{H} = \mathrm{rot}\boldsymbol{A}$. It is crucial to note that the vector potential should be introduced in the Hamiltonian in the LK representation not in the Bloch bands, since the dependences on the wave vector, $\bm{k}$, are explicitly and correctly represented only in this representation. The energy spectra in clean systems in the presence of magnetic field are known to be $E_N = (\sqrt{2}v/l)\sqrt{|N|}\mathrm{sgn}(N)$ with integer $N$ and $l = \sqrt{c/eH}$. In actual systems, however, there exists always disorder resulting in finite damping of the electronic spectra and the evaluation of physical properties can be very intricate in the presence of both disorder and magnetic field. The consequences of such disorder may vary from case to case as has been clarified by Shon and Ando[20] for the transport properties. In the present paper we assume for the sake of simplicity and clarity a damping, $\Gamma$, which is independent of energy, so that analytical results are obtained and then the anomalous contributions of inter-band effects of magnetic field are clarified.

The orbital susceptibility, $\chi$, in the limit of weak magnetic field is calculated based on the formula given in ref. 16, which is exact for any Bloch electrons. (Here the gauge invariance relative to the vector potential has been explicitly demonstrated.) Application of this formula to the present model Hamiltonian, eq. (1), results in as follows,

$$\chi = \frac{e^2}{c^2} T \sum_n \sum_{\bm{k}} \mathrm{Tr} G\gamma_x G\gamma_y G\gamma_x G\gamma_y \tag{2}$$

where $\gamma_\alpha = v\sigma_\alpha$ and $G = G(\bm{k}, \mathrm{i}\tilde{\varepsilon}_n) = [\mathrm{i}\tilde{\varepsilon}_n - \mathcal{H}]^{-1}$ is the thermal Green function with $\mathrm{i}\tilde{\varepsilon}_n = \mathrm{i}\varepsilon_n + \mathrm{i}\Gamma\mathrm{sgn}(\varepsilon_n)$ ($\varepsilon_n = (2n+1)\pi T$). It is straightforward to see that eq. (2) at absolute zero leads to a simple result as given as follows per unit area (including the spin degeneracy factor, 2, hereafter),

$$\chi = -4\frac{e^2}{c^2}v^4 T \sum_n \sum_{\bm{k}} \frac{(\mathrm{i}\tilde{\varepsilon}_n)^2[(\mathrm{i}\tilde{\varepsilon}_n)^2 - 2v^2k^2]}{[(\mathrm{i}\tilde{\varepsilon}_n)^2 - v^2k^2]^4} \tag{3a}$$

$$= \frac{2}{3}\frac{e^2v^2}{\pi^2 c^2}\Gamma \int_{-\infty}^{\infty} \mathrm{d}\varepsilon f(\varepsilon) \frac{\varepsilon}{(\varepsilon^2 + \Gamma^2)^2} \tag{3b}$$

$$= -\left(\frac{e^2v^2}{3\pi^2 c^2 \Gamma}\right) \frac{1}{1+X^2} \equiv \chi_0 \frac{1}{1+X^2} \tag{3c}$$

where $f(\varepsilon)$ is the Fermi distribution function and $X = \mu/\Gamma$. This result at absolute zero with finite damping is in accordance with the results of McClure[8] in clean systems at finite temperature.

The Hall conductivity, $\sigma_{xy}$, in the weak field limit will be evaluated by the general formalism given in ref. 17 where the finite Fourier component, $\bm{q}$, has been introduced in the vector potential $\bm{A}(\bm{r}) = \bm{A}_{\bm{q}}\mathrm{e}^{\mathrm{i}\bm{q}\cdot\bm{r}}$ in order to reach the gauge-invariant result explicitly. In the present Hamiltonian, eq. (1), one starts with processes shown in Fig. 2 to the linear order of $\bm{A}_{\bm{q}}$ where $\bm{k}_\pm = \bm{k} \pm \bm{q}/2$, $\mathrm{i}\tilde{\varepsilon}_{n-} = \mathrm{i}\tilde{\varepsilon}_n - \mathrm{i}\omega_\lambda$ and extracts the contribution linear in $\bm{q}$ resulting

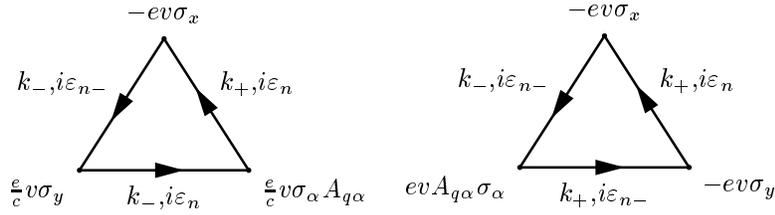

Fig. 2.　Processes for Hall conductivity, where $\sigma_\alpha$ ($\alpha = x, y$) is the pauli matrix and $\bm{k}_\pm = \bm{k} \pm \bm{q}/2$, $i\varepsilon_{n-} = i\varepsilon_n - i\omega_\lambda$, $i\omega_\lambda$ being the external frequency.

in

$$\sigma_{xy} = \frac{4e^3v^4}{i\omega c}(q_x A_{\bm{q}y} - q_y A_{\bm{q}x})T\sum_n \sum_{\bm{k}}[F_2(i\varepsilon_{n-})F_1(i\varepsilon_n) - F_1(i\varepsilon_{n-})F_2(i\varepsilon_n)] \tag{4}$$

where $F_l(i\varepsilon_n) = i\tilde{\varepsilon}_n/[(i\tilde{\varepsilon}_n)^2 - v^2k^2]^l$, ($l = 1, 2$). Here $i(q_x A_{\bm{q}y} - q_y A_{\bm{q}x}) = H$ is to be noted. By summing over thermal frequency and analytically continuing, $i\omega_\lambda \to \omega$, one ends with

$$\sigma_{xy} = -i\frac{2e^3v^4 H}{\pi c}\sum_{\bm{k}}\int d\varepsilon\left[2f(\varepsilon)\left\{\frac{(\varepsilon + i\Gamma)^3}{[(\varepsilon + i\Gamma)^2 - v^2k^2]^4} - \text{c.c.}\right\}\right.$$
$$\left.+ f'(\varepsilon)\left\{\frac{(\varepsilon - i\Gamma)}{[(\varepsilon - i\Gamma)^2 - v^2k^2]^2}\frac{(\varepsilon + i\Gamma)}{[(\varepsilon + i\Gamma)^2 - v^2k^2]} - \text{c.c.}\right\}\right] \tag{5a}$$

$$= \frac{e^3v^2 H}{4\pi^2 c\Gamma^2}\left[\frac{1-X^2}{1+X^2} - \frac{1+X^2}{2|X|}\Phi(X) - \frac{8}{3}\frac{X^2}{(1+X^2)^2}\right]\bigg/X \tag{5b}$$

$$\equiv \sigma_{xy}^0 K_{xy}(X) \tag{5c}$$

where $\Phi(X) = \pi/2 - \tan^{-1}[(1-X^2)/2|X|]$ and $\sigma_{xy}^0 = e^3v^2H/4\pi^2c\Gamma^2$.

We note that $K_{xy} \sim -\frac{16}{3}X$ for $|X| \leq 1$ and $K \sim -\frac{\pi}{2}\text{sgn}(X)$ for $|X| \gg 1$. In order to see the physical implication of the result of $\sigma_{xy}$, diagonal conductivity, $\sigma$, is evaluated as follows,

$$\sigma = \frac{e^2}{2\pi^2}\left[1 + \frac{1+X^2}{2|X|}\Phi(X)\right] \equiv \frac{e^2}{2\pi^2}K_{xx}. \tag{6}$$

We note $\sigma = e^2/\pi^2$ at $\mu = 0$, which is the same as that of Shon and Ando[20] in the case of graphene for short range impurity scattering and similar to that of Katayama et al.[22] for $\alpha$-ET$_2$I$_3$ with different energy spectra. On the other hand, $\sigma \sim (e^2/4\pi)|\mu|/\Gamma$ for $|X| \gg 1$. By use of eqs. (5c) and (6) the Hall coefficient $R = \sigma_{xy}/H\sigma^2 = \frac{\pi^2v^2}{ec\Gamma^2}\frac{K_{xy}}{K_{xx}^2} \equiv R_0\frac{K_{xy}}{K_{xx}^2}$ can be estimated from which effective carrier density could be deduced. However, in contrast to semiconductors with finite band gap, the meaning of "carrier density" will not be unique in the present case since the group velocity is finite at zero-gap region. It is interesting to note that $R$ in the present case is vanishingly small as $X \to 0$ indicating that "the effective carrier density" is diverging, while it takes following asymptotic values as $|X| \gg 1$,

$$\frac{1}{ecR} \to -2\int_0^\mu d\varepsilon N(\varepsilon) = -\frac{\mu^2}{2\pi v^2}\text{sgn}(\mu) \tag{7}$$

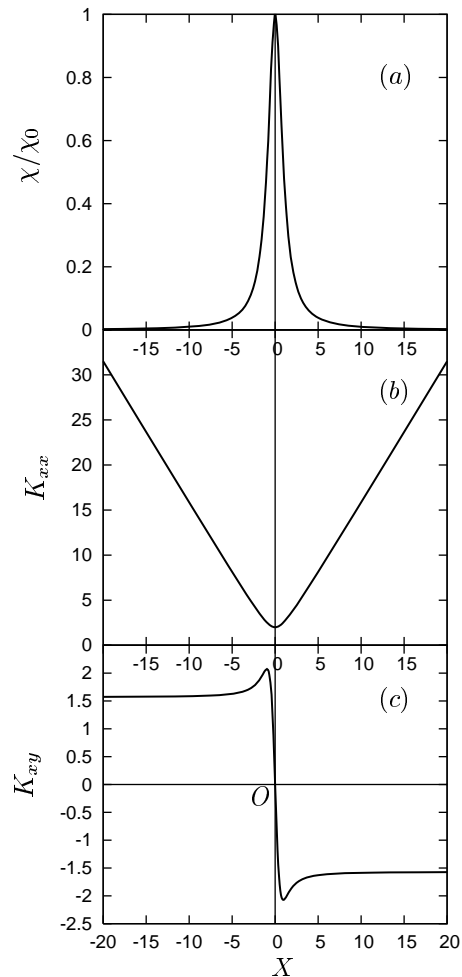

Fig. 3. The dependences on the Fermi energy, $\mu$, relative to the damping, $\Gamma$, ($X = \mu/\Gamma$) of the orbital diamagnetism (a), conductivity (b) and the Hall conductivity (c).

where $N(\varepsilon) = |\varepsilon|/2\pi v^2$ is the density of states per spin in clean systems.

Summarizing the analytical results so far obtained, in Fig. 3, are shown the dependences on the Fermi energy, $\mu$, of the orbital susceptibility, $\chi$, eq. (3c), the conductivity, eq. (6) and the Hall conductivity, eq. (5c). In Fig. 4, the Hall coefficient $R/R_0$ and its inverse $(R/R_0)^{-1}$ proportional to effective carrier density are shown. It is to be noted that, if the energy dependences of $\Gamma$ is present and $\Gamma \sim \varepsilon$, the conductivity will be essentially independent of the Fermi energy.[20]

Extension of the present studies to $\alpha$-ET$_2$I$_3$, which is described by the "tilted Weyl equation",[21,22] and Bi-Sb alloys with strong spin-orbit interaction are interesting targets to be studied.

Author thanks for M. Nakamura for discussions, and T. Suzuki and T. Aonuma for technical help. This work is financially supported by Grant-in-Aid for Scientific Research on Priority Area"Invention of anomalous quantum materials -New physics through innovative materials-"

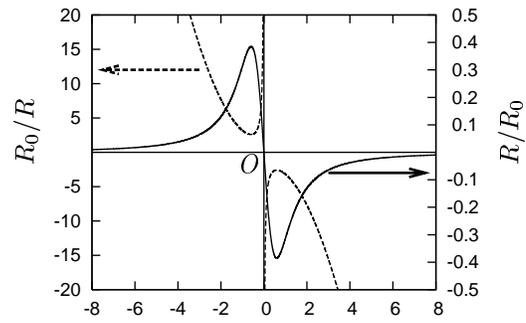

Fig. 4. The Fermi energy dependences of the Hall coefficient, $R$ (solid line), and the effective carrier number $R^{-1}$ (dashed line).

of Ministry of Education, Culture, Sports, Science and Technology, Japan.

J. Phys. Soc. Jpn.